\documentclass[aps,notitlepage,nofootinbib,10pt,prd,superscriptaddress,preprintnumbers]{revtex4-1}
\usepackage{latexsym}
\usepackage{amsfonts}
\usepackage{amssymb}
\usepackage{amsmath}
\usepackage{xhfill}

\usepackage{color}

\def\mpl{M_{\rm Pl}}
\def\K{{\cal K}}
\def\d{\mathrm{d}}
\def\ta{}
\def\ta{{\zeta_a}}

\newcommand{\eff}{{\rm eff}}

\begin{document}
\preprint{YITP-15-124}
\preprint{Imperial/TP/2016/AEG/1}
\preprint{IPMU15-0222}

\title{New Quasidilaton theory in Partially Constrained Vielbein Formalism}
\date\today

\author{Antonio De Felice}
\email{antonio.defelice@yukawa.kyoto-u.ac.jp}
\affiliation{Yukawa Institute for Theoretical Physics, Kyoto University, 606-8502,
Kyoto, Japan}

\author{A. Emir G\"umr\"uk\c c\"uo\u glu}
\email{a.gumrukcuoglu@imperial.ac.uk}
\affiliation{
Theoretical Physics Group, Blackett Laboratory,
Imperial College London, South Kensington Campus, London, SW7 2AZ,
UK}

\author{Lavinia Heisenberg}
\email{lavinia.heisenberg@eth-its.ethz.ch}
\affiliation{Institute for Theoretical Studies, ETH Zurich, Clausiusstrasse 47, 8092 Zurich, Switzerland}

\author{Shinji Mukohyama}
\email{shinji.mukohyama@yukawa.kyoto-u.ac.jp}
\affiliation{Yukawa Institute for Theoretical Physics, Kyoto University, 606-8502, Kyoto, Japan}
\affiliation{Kavli Institute for the Physics and Mathematics of the Universe (WPI), The University of Tokyo Institutes for Advanced Study, The University of Tokyo, Kashiwa, Chiba 277-8583, Japan}

\author{Norihiro Tanahashi}
\email{N.Tanahashi@damtp.cam.ac.uk}
\affiliation{Department of Applied Mathematics and Theoretical Physics, Centre for Mathematical Sciences, University of Cambridge, Wilberforce Road, Cambridge CB3 0WA, UK}

\begin{abstract}
In this work we study the partially constrained vielbein formulation of the new quasidilaton theory of massive gravity which couples
to both physical and fiducial metrics simultaneously via a composite effective metric. This formalism improves the new quasidilaton 
model since the Boulware-Deser ghost is removed fully non-linearly at all scales. This also yields crucial implications in the cosmological applications. 
We derive the governing cosmological background evolution and study the stability of the attractor solution.
\end{abstract}

\maketitle

\section{Introduction}

Introduction of a non-zero graviton mass is one of the simplest modifications to the general theory of relativity.
Massive gravity theories were first proposed by Fierz and Pauli in 1939 \cite{Fierz:1939ix}, and it gained renewed interest since the discovery of the nonlinear completion by de Rham, Gabadadze and Tolley (dRGT) \cite{deRham:2010kj}, in which the Bouleware-Deser (BD) ghost \cite{BoulwareD} is absent even at the non-linear level \cite{Hassan12b}. The theory is elaborate but encloses indispensable promising properties, the technically naturalness being one of them \cite{deRham:2012ew,deRham:2013qqa,Heisenberg:2014rka}.\footnote{See Ref.~\cite{deRham:2014zqa} for a detailed review.} 
Since its inception the allowed cosmology of dRGT theory has been widely discussed \cite{deRham:2010tw,Koyama:2011xz,Koyama:2011yg,deRham:2011by,PhysRevD.84.124046,Koyama:2011wx,Volkov:2011an,Comelli:2011zm,Gumrukcuoglu:2011zh,Volkov:2012cf,DeFelice:2013awa}.
This topic is attracting interest because a nonzero graviton mass may induce the accelerated expansion of the universe and could be an alternative to the cosmological constant and dark energy. The original form of dRGT massive gravity theory encountered several problems such as absence of solutions describing a realistic cosmology without pathological behaviors
\cite{Gumrukcuoglu:2011zh,PhysRevLett.109.171101,Fasiello:2012rw,Langlois:2012hk,MartinMoruno:2013gq}, and many extensions of the theories were proposed to alleviate these issues \cite{Huang:2012pe,PhysRevD.87.064037,DeFelice:2013tsa,deRham:2014gla,DeFelice:2015yha}.

In this work, we focus on a specific extension, obtained via a scalar field associated with a global \emph{quasi-dilaton} symmetry \cite{PhysRevD.87.064037}
\begin{equation}
\sigma \to \sigma+\sigma_0\,,\qquad 
f_{\mu\nu}\to {\rm e}^{-2\,\sigma_0/\mpl}f_{\mu\nu}\,,
\label{eq:quasidilatonsymmetry}
\end{equation}
where $f_{\mu\nu}$ corresponds to the non-dynamical \emph{fiducial metric}, while the \emph{physical metric} $g_{\mu\nu}$ is invariant under these transformations. The action invariant under this symmetry depends on the combination ${\rm e}^{2\,\sigma/\mpl}\,f_{\mu\nu}$, where the conformal factor allows the otherwise absent flat cosmological solutions with self-accelerated expansion. On the other hand, the original theory has also pathological cosmological solutions \cite{Gumrukcuoglu:2013nza,D'Amico:2013kya} although generalization of the original action to include an additional coupling constant does address this issue \cite{DeFelice:2013tsa,DeFelice:2013dua}. Even if the background evolution is insensitive to the presence of this new coupling constant, the stability of the perturbations crucially depends on it, hence allowing to cure the reported instability in the original formulation \cite{DeFelice:2013tsa,DeFelice:2013dua,Heisenberg:2015voa}. 
In the standard formulation of massive gravity, the graviton mass is at most of the the same order as the Hubble expansion rate today. Since the effective mass of the tensor perturbations is proportional to the graviton mass, this enforces the effective mass of gravitational waves to be of the same order. In the presence of the new coupling constant in the extended quasidilaton theory this restriction can be softened and it could even accommodate an explanation for the large-angle suppression of power in the microwave background \cite{Kahniashvili:2014wua}. 

Another promising extension is the new quasidilaton theory which is based on modifying the coupling of the quasidilaton field to the metrics~\cite{Mukohyama:2014rca}.
In this case, instead of minimally coupling the quasidilaton Lagrangian only to the physical metric, it now couples to a combination of both physical and fiducial metrics. 
Such a composite coupling generically revives the BD ghost into the theory \cite{deRham:2014naa,deRham:2014fha}, nevertheless there is a unique effective metric in the sense that it maintains the theory ghost free up to the strong coupling scale \cite{deRham:2015cha,Huang:2015yga,Heisenberg:2015iqa}. This modification also introduces an additional parameter to the theory which can be tuned to make the accelerated expansion and stability of cosmological solutions compatible with each other. One downside of this theory was that the additional parameter in the quasidilaton coupling must be tuned to make the mass of the ghost degree of freedom sufficiently high.
Such a tuning becomes unnecessary if the ghost degree of freedom is absent at all scales. One way to realize this was initially thought to come from redefining the theory in the unconstrained vielbein formulation, in which the theory is formulated in the vielbein language and the dynamics of the vielbein is determined by its own equations of motion but not by the symmetric vielbein conditions \cite{Hinterbichler:2015yaa}. Unfortunately, in this formalism the ghost degree of freedom is reintroduced to the theory when the rotation part of the vielbein is integrated out \cite{deRham:2015cha}. Instead, this issue can be circumvented by adopting the \emph{partially constrained vielbein formalism} \cite{DeFelice:2015yha} where the rotational part of the vielbein is symmetric by construction and the boost part is determined by their equations of motion. In this formalism the ghost degree of freedom is absent fully non-linearly, hence the allowed parameter region is enlarged as described above.

In this paper we examine de Sitter solutions of new quasidilaton theory in the partially constrained vielbein formalism and provide a stability analysis of perturbations.
The background dynamics and the dispersion relations for the tensor and vector perturbations in this extended theory is the same as in the metric formulation, while the crucial difference arises in the scalar perturbations. 
After introducing the partially constrained vielbein formalism in Section ~\ref{Sec:partially}, we derive the late--time de Sitter attractor background in Section~\ref{sec:bgsol}. Section ~\ref{sec:pert} is devoted to the stability conditions against ghost and gradient instabilities for this background. We conclude this work with summary and discussions in Section~\ref{Sec:summary}. The paper is supplemented by the Appendix, where the results for the metric formulation is summarized for comparison.

%%%%%%%%%%%%%%%%%%%%%%%%%%%%%%%%%%%%%%%%%%%%%%%%
%%%%%%%%%%%%%%%%%%%%%%%%%%%%%%%%%%%%%%%%%%%%%%%%
%%%%%%%%%%%%%%%%%%%%%%%%%%%%%%%%%%%%%%%%%%%%%%%%
\section{Partially constrained formulation of New quasidilaton theory}
\label{Sec:partially}

In this section we would like to introduce the new quasidilaton theory in the partially constrained vielbein formulation. We shall adapt the vielbein formulation of massive gravity. For this purpose, we express the two metrics by the following vielbeins as 
\begin{equation}
 g_{\mu\nu}  =  \eta_{\cal AB} e^{\cal A}{}_{\mu} e^{\cal B}{}_{\nu}  \qquad \text{and} \qquad
 f_{\mu\nu}  =  \eta_{\cal AB} E^{\cal A}{}_{\mu} E^{\cal B}{}_{\nu}\,,
\label{eq:metrics}
\end{equation}
where ${\cal A,B}=0,1,2,3$ are the indices in the orthonormal bases.
Since in quasi-dilaton theories, the fiducial metric only appears with the conformal factor ${\rm e}^{-2\,\sigma/\mpl}$, it is useful to further define an orthonormal basis for this combination as
\begin{equation}
{\rm e}^{-2\,\sigma/\mpl}f_{\mu\nu} = \eta_{\cal AB} \tilde{E}^{\cal A}{}_{\mu} \tilde{E}^{\cal B}{}_{\nu}\,, \quad \tilde{E}^{\cal A}{}_{\mu} = {\rm e}^{-\,\sigma/\mpl}E^{\cal A}{}_{\mu}\,.
\end{equation}
In this formulation, the mass term corresponds to the interactions between the two vielbeins $e^{\cal A}_{\;\mu}$ and $\tilde{E}^{\cal A}_{\;\mu}$, constructed out of their wedge products \cite{Hinterbichler:2012cn}
\begin{eqnarray}
 {\cal L}_0 & = & \frac{1}{24}
  \epsilon^{\mu\nu\rho\sigma}\epsilon_{\cal ABCD}
  \tilde{E}^{\cal A}{}_{\mu}\tilde{E}^{\cal B}{}_{\nu}\tilde{E}^{\cal C}{}_{\rho}\tilde{E}^{\cal D}{}_{\sigma}\,,
  \nonumber\\
 {\cal L}_1 & = & \frac{1}{6} 
  \epsilon^{\mu\nu\rho\sigma}\epsilon_{\cal ABCD}
  \tilde{E}^{\cal A}{}_{\mu}\tilde{E}^{\cal B}{}_{\nu}\tilde{E}^{\cal C}{}_{\rho}e^{\cal D}{}_{\sigma}\,,
  \nonumber\\
 {\cal L}_2 & = & \frac{1}{4}
  \epsilon^{\mu\nu\rho\sigma}\epsilon_{\cal ABCD}
  \tilde{E}^{\cal A}{}_{\mu}\tilde{E}^{\cal B}{}_{\nu}e^{\cal C}{}_{\rho}e^{\cal D}{}_{\sigma}\,,
  \nonumber\\
 {\cal L}_3 & = & \frac{1}{6}
  \epsilon^{\mu\nu\rho\sigma}\epsilon_{\cal ABCD}
  \tilde{E}^{\cal A}{}_{\mu}e^{\cal B}{}_{\nu}e^{\cal C}{}_{\rho}e^{\cal D}{}_{\sigma}\,,
  \nonumber\\
 {\cal L}_4 & = & \frac{1}{24}
  \epsilon^{\mu\nu\rho\sigma}\epsilon_{\cal ABCD}
  e^{\cal A}{}_{\mu}e^{\cal B}{}_{\nu}e^{\cal C}{}_{\rho}e^{\cal D}{}_{\sigma}\,,
  \label{eqn:def-gravitonmassterm}
 \end{eqnarray}
where the Levi-Civita symbols are normalized as $\epsilon_{0123}=1=-\epsilon^{0123}$.
The dual bases of the vielbeins are defined such that
 \begin{eqnarray}
 &&E^{\cal A}{}_{\mu}E_{\cal A}{}^{\nu} = \delta_{\mu}^{\nu}, \quad E^{\cal A}{}_{\mu} E_{\cal B}{}^{\mu} = \delta^{\cal A}_{\cal B} \nonumber\\
 &&e^{\cal A}{}_{\mu}e_{\cal A}{}^{\nu} = \delta_{\mu}^{\nu}, \quad e^{\cal A}{}_{\mu}e_{\cal B}{}^{\mu} = \delta^{\cal A}_{\cal B}\,.
\end{eqnarray}

The invariance under the overall local Lorentz transformation of the two vielbeins allows us to 
fix the gauge freedom associated with the boost part of the overall local Lorentz transformation. We can do that by choosing the Arnowitt-Deser-Misner (ADM) form for the fiducial vielbein
\begin{equation}
 E^{\mathcal{A}}_{\ \mu} = 
  \left(
   \begin{array}{cc}
    M & 0\\
    M^kE^I_{\ k} & E^I_{\ j}
    \end{array}
       \right)\,, \label{eqn:ADMvielbein}
\end{equation}
where $I,J = 1,2,3$ correspond to the spatial indices in the orthonormal basis. 
Here, $M$, $M^i$ and $E^I_{\ j}$ are the fiducial lapse, the fiducial shift and the fiducial spatial vielbein since the corresponding $4$-dimensional fiducial metric $f_{\mu\nu}$ becomes
\begin{equation}
 f_{\mu\nu}dx^{\mu}dx^{\nu} = -M^2dt^2 + f_{ij}(dx^i+M^idt)(dx^j+M^jdt), \quad
  f_{ij} = \delta_{IJ}E^I_{\ i}E^J_{\ j}. 
\end{equation}
We cannot bring the physical vielbein into the ADM form simultaneously, however we can write it in the boosted ADM form, such as
\begin{equation}
e^{\cal A}_{~~\mu} = \left(e^{-\omega}\right)^{\cal A}_{~~\cal B}\,\varepsilon^{\cal B}_{~~\mu}\,,
\end{equation}
where we have introduced the ADM vielbein
\begin{equation}
\varepsilon^{\cal A}_{~~\mu} = \left(
\begin{array}{ll}
N & 0_j\\
\varepsilon^I_{~k}N^k & \varepsilon^I_{~j}
\end{array}
\right)\,,
\label{eq:ADMdecomp}
\end{equation}
and $\left(e^{-\omega}\right)^{\cal A}_{~~\cal B}$ represents a
general proper boost transformation. The ADM vielbein is defined
through the physical lapse $N$, the physical shift $N^i$, and the
physical spatial vielbein $e^I_{\ j}$, whereas the Lorentz-boost
transformation is a function of the boost parameter $b_I$, as in
$\omega^{\mathcal A}{}_{\mathcal B}=\sum_{I}b_I ({L_I})^{\mathcal
  A}{}_{\mathcal B}$,
and $L_I$ are the three generators of the boost. In terms of these
variables the physical metric becomes
\begin{equation}
 g_{\mu\nu}dx^{\mu}dx^{\nu} = -N^2dt^2 + g_{ij}(dx^i+N^idt)(dx^j+N^jdt), \quad
  g_{ij} = \delta_{IJ}e^I_{\ i}e^J_{\ j}. 
\end{equation}

Besides fixing the boost part of the overall local Lorentz transformation as (\ref{eqn:ADMvielbein}), we further impose the following symmetric condition on the $3$-dimensional spatial vielbein \cite{DeFelice:2015yha}
\begin{equation}
 Y_{IJ} = Y_{JI}, \label{eqn:YIJ=YJI}
\end{equation}
where $Y_{IJ} \equiv E_I^{\ l}\delta_{JK}e^K_{\ l}$. This condition (\ref{eqn:YIJ=YJI}) does not correspond to a gauge condition but is rather a physical condition yielding a different formulation of the original theory, which was dubbed ``partially constrained vielbein formulation''. This physical condition (\ref{eqn:YIJ=YJI}) puts the spatial components and the temporal component of the vielbein on different footing, violating local Lorentz invariance in the gravity sector. 

The composite metric is constructed as \cite{deRham:2014naa,Noller:2014sta}
\begin{equation}
 g^{\rm eff}_{\mu\nu} =
  \eta_{\cal AB} e^{\cal A}_{\rm eff}{}_{\mu} e^{\cal B}_{\rm eff}{}_{\nu}\,,
\label{eq:geff-def}
\end{equation}
where the composite vielbein is a linear combination of the two vielbeins, enriched with the quasidilaton field
\begin{equation}
 e^{\cal A}_{\rm eff}{}_{\mu} =
  \alpha e^{\cal A}{}_{\mu} + \beta \tilde{E}^{\cal A}{}_{\mu}
=    \alpha e^{\cal A}{}_{\mu} + \beta e^{\sigma/\mpl}  E^{\cal A}{}_{\mu}\,.
\label{eq:tetradeff-def}
\end{equation}
Finally, the general action consisting of the dynamical vielbein, the quasidilaton field that lives on the effective vielbein and the standard matter fields living on the dynamical vielbein of the theory reads
\begin{equation}
S = \frac{M_p^2}{2}\int d^4x \left[\det{e}\,R[e] + 2\,m^2\sum_{n=0}^4 \beta_n\,{\cal L}_n\right] +\int d^4x \det{e_\eff}\,{\cal L}_\sigma(g_\eff, \partial_\mu\sigma)+\int d^4x \det{e}\, {\cal L}_{\rm matter}\,.\label{eq:action}
\end{equation}
The most general Lagrangian for the quasi-dilaton field that is invariant under (\ref{eq:quasidilatonsymmetry}) was given in Ref.~\cite{DeFelice:2013dua}. In this paper, for the sake of simplicity, we choose the canonical action
\begin{equation}
{\cal L}_\sigma = -\frac{\omega}{2}\,g_\eff^{\mu\nu}\partial_\mu\sigma\partial_\nu\sigma\,.
\end{equation}

%%%%%%%%%%%%%%%%%%%%%%%%%%%%%%%%%%%%%%%%%%%%%%%%
%%%%%%%%%%%%%%%%%%%%%%%%%%%%%%%%%%%%%%%%%%%%%%%%
%%%%%%%%%%%%%%%%%%%%%%%%%%%%%%%%%%%%%%%%%%%%%%%%

\section{Cosmological background and late time de Sitter solution} \label{sec:bgsol}
We shall as next study the cosmological background evolution of the quasidilaton living on the composite effective metric\footnote{Further cosmological implications of the composite effective metric have been studied in \cite{deRham:2014naa,Enander:2014xga,Gumrukcuoglu:2014xba,Solomon:2014iwa,Gao:2014xaa,Gumrukcuoglu:2015nua,Heisenberg:2015wja,Lagos:2015sya} and dark matter phenomenology in \cite{Blanchet:2015sra,Blanchet:2015bia,Bernard:2015gwa}.}. First of all, we will choose the unitary gauge, and we will consider the class of theories defined by the following fiducial vielbein components as in
\begin{equation}
M=M(t)\,,\qquad M^i=0\,,\qquad E^I{}_j=a_0\,\delta^I{}_j\,,
\end{equation}
so that the fiducial metric is Minkowski with the general lapse function $M(t)$, $f_{\mu\nu}dx^{\mu}dx^{\nu}=-M^2(t)dt^2+a_0^2\delta_{ij}dx^idx^j$. 
Furthermore, we will assume that on the background $b_I=0$, such that the symmetry condition for $Y_{IJ}$ implies that the 3d ADM vielbein on the background $\varepsilon^I{}_j(t)$ is symmetric, allowing us to fix the latter as 
\begin{equation}
\varepsilon^I{}_j(t)=a(t)\, \delta^I{}_j\,.
\end{equation}
In the following we will also fix, without loss of generality (i.e.\ keeping the time-like St\"uckelberg field, and thus the fiducial lapse, as a general function of the time $t$), $N(t)=1$. 
The corresponding lapse function and scale factor of the background composite effective vielbein then become 
\begin{equation}
N_{\rm eff}= \alpha+\beta r X\,, \qquad a_{\rm eff}=(\alpha+\beta X)\, a\,,
\end{equation}
respectively, where we defined
\begin{equation}
X\equiv \frac{a_0 e^{\sigma/\mpl}}{a}\,,\qquad
r\equiv \frac{a\,M}{a_0}\,.
\end{equation}
A quasidilaton configuration compatible with this setup is a homogeneous one, i.e.\ $\sigma=\sigma(t)$. Finally, since we will focus on de Sitter attractor solutions, in the remainder of the text, we consider only a cosmological constant in the matter sector that only couples to the physical vielbein.
The first trivial observation is that the background equations of motion in the partially constrained vielbein formulation are exactly the same as in the metric formulation. This was already pointed out in \cite{DeFelice:2015yha} for the case of dRGT massive gravity. 
The equations of motion for this configuration is obtained as\footnote{For the details of the background equations, we refer the reader to Appendix \ref{Sec:metric} where these are derived for the metric formulation. As stressed in the main text, the two formulations coincide at the background level.}
\begin{align}
({\bf i})\;& 3\,H^2 = \Lambda+m^2\rho_{m,g} +\frac{\alpha\,\omega\,a_\eff^3\dot{\sigma}^2}{2\,\mpl^2a^3N_\eff^2}\,,\nonumber\\
({\bf ii})\;& 2\,\dot{H}= m^2 J\,X\,(r-1)-
\frac{\alpha\,\omega\,a_\eff^3\left(1+\frac{a\,N_\eff}{a_\eff}\right)\dot{\sigma}^2}{2\,\mpl^2a^3N_\eff^2}
\,,\nonumber\\
({\bf iii})\;& \frac{1}{N_\eff}\,\partial_t\,\left(\frac{\dot{\sigma}}{N_\eff}\right)+3\,H_\eff\frac{\dot{\sigma}}{N_\eff}
+
\frac{\beta\,X\,\left(\frac{r}{N_\eff}-\frac{3\,a}{a_\eff}\right)\dot{\sigma}^2}{2\,\mpl\,N_\eff^2} + \frac{m^2\mpl a^3 X}{\omega\,a_\eff^3 N_\eff}\left[4\,r\,X^3 \rho_{m,f}-3\,J\,(r-1)\right]=0\,,\nonumber\\
({\bf iv})\;&\frac{\beta\,\omega\,a_\eff^3X\,\dot{\sigma}^2}{2\,a^3\,N_\eff^2}+m^2\mpl^2 X^4 \rho_{m,f} = \left(\frac{a_0}{a}\right)^4\,\kappa\,,
\label{eq:eomsBG}
\end{align}
where $\kappa$ is a free integration constant and we defined the expansion rate of the composite effective cosmology as 
\begin{equation}
H_{\eff} \equiv \frac{\dot{a}_\eff}{a_\eff N_\eff} = \frac{a}{a_\eff N_\eff}\,\left(\alpha\,H+\beta\,X\,\frac{\dot{\sigma}}{\mpl}\right)\,.
\label{eq:heff}
\end{equation}
In Eq.~(\ref{eq:eomsBG}) and below, instead of using the $\beta_n$ coefficients, we employ the following polynomial
\begin{equation}
U(X)=\beta_0X^4 +4\,\beta_1X^3+6\,\beta_2X^2+4\,\beta_3X+\beta_4\,,
\end{equation}
which allows us to define%
\footnote{We remark that the four functions defined in Eq.~(\ref{eq:Xfunctionsbeta}) are not enough to solve for the five parameters $\beta_n$. 
The fifth combination of $\beta_n$ can be absorbed into the cosmological constant $\Lambda\equiv m^2 \beta_4$.
}
\begin{eqnarray}
\rho_{m,g}(X)&\equiv& U(X)-\frac{X}{4}\,U'(X)\,,\nonumber\\
\rho_{m,f}(X)&\equiv& \frac{1}{4\,X^3}\,U'(X)\,,\nonumber\\
J(X)&\equiv& \frac{1}{3}\,\rho_{m,g}'(X)\,,\nonumber\\
\Gamma(X,r)&\equiv& X\,J(X)+\frac{X^2(r-1)}{2}\,J'(X)\,.
\label{eq:Xfunctionsbeta}
\end{eqnarray}

We now look for de Sitter solutions, i.e.\ $H=H_0={\rm constant}$. Combining (i) and (iv) in Eq.~(\ref{eq:eomsBG}), we find:
\begin{equation}
3\,H_0^2 = \Lambda +m^2\left(\rho_{m,g}-\frac{\alpha\,X^3}{\beta} \rho_{m,f}\right) + \frac{a_0^4}{a^4}\,\frac{\alpha\,\kappa}{\mpl^2\beta\,X}\,.
\label{eq:eq-i-iv}
\end{equation}
The last term in the right hand side of the above equation redshifts as $a^{-4}$, while the part
$\propto m^2$ consists of terms which at most redshift as
$X^3\sim a^{-3}$. Therefore, for a late time solution, we can safely ignore the
term arising from the integration constant $\kappa$. In this late time
limit, Eq.~(\ref{eq:eq-i-iv}) thus implies that $X=X_0={\rm constant}$ on the attractor solution.

Following a similar argument, we can also combine (ii)
and~(iv) in Eq.~(\ref{eq:eomsBG}). As $X$ is constant, the functions $J(X_0)$ and $\rho_{m,f}(X_0)$ are also
constant on the attractor, implying that $r=r_0={\rm constant}$ on the
late time de Sitter solution. In other words, both $a_\eff/a$ and
$N_\eff$ are constant. Then, Eq.~(\ref{eq:eomsBG}.iv) trivially implies that
$\dot{\sigma}/N_\eff = {\rm constant}$. Finally, using all this
information in Eq.~(\ref{eq:eomsBG}.iii) above, while simultaneously using Eqs.~(ii) and
(iv), one can show that
\begin{equation}
 \dot{\sigma}\,\left(\frac{\dot{\sigma}}{\mpl}-H_0\right)=0\,.
\end{equation}
The non-trivial solution gives the background evolution of the quasi-dilaton as
\begin{equation}
\frac{\dot{\sigma}}{\mpl} = H_0\,.
\label{eq:sigmadotattractor}
\end{equation}

In the following, the strategy for going on shell is as follows. We
solve (i) for $\Lambda$, (ii) and (iv) for $J(X_0)$ and
$\rho_{m,f}(X_0)$. Eq.~(iii) is trivially satisfied once the solution
(\ref{eq:sigmadotattractor}) is used. Then, on the attractor, the derivatives
of equations (\ref{eq:eomsBG}) are automatically satisfied.

For the sake of a clear notation, in the remainder of the text we omit the subscript $0$ denoting the values on the attractor. 

%%%%%%%%%%%%%%%%%%%%%%%%%%%%%%%%%%%%%%%%%%%%%%%%
%%%%%%%%%%%%%%%%%%%%%%%%%%%%%%%%%%%%%%%%%%%%%%%%
%%%%%%%%%%%%%%%%%%%%%%%%%%%%%%%%%%%%%%%%%%%%%%%%
\section{Stability of the de Sitter attractor}\label{sec:pert}

In this section we will study the stability of the attractor solution. Then we need to introduce perturbation variables for all the dynamical variables. As for the physical ADM vielbein, we can write it
\begin{equation}
\varepsilon^{\cal A}_{~~\mu} = \left(
\begin{array}{ll}
1+\Phi & 0_j\\
\varepsilon^I_{~k}N^k & \varepsilon^I_{~j}
\end{array}
\right)\,,
\label{eq:ADMdecomp-pert}
\end{equation}
where the shift vector is perturbed as
\begin{equation}
N_i = a\,(B_k+\partial_k B)\,,
\label{eq:decompN}
\end{equation}
whereas the three dimensional ADM vielbein as
\begin{equation}
\varepsilon^I_{~i} = a\,(1+\,\psi)\delta^I_i+\frac{a\,\delta^{Ij}}{2}\left[\gamma_{ij}+\partial_{(i}E_{j)}+\left(\delta^k_i\delta^l_j-\frac{1}{3}\,\delta_{ij}\delta^{kl}\right)\partial_k\partial_lE\right]\,,
\label{eq:decompe}
\end{equation}
where
$\delta^{ij}\partial_iB_j=\delta^{ij}\partial_iE_j=\delta^{ij}\partial_i\gamma_{jk}
= \delta^{ij}\gamma_{ij}=0$. Recall that we are performing the perturbation analysis above a background where the
Lorentz transformation exactly vanishes
$\omega^{\cal A}_{~~\cal B}(t)=0$.  Furthermore, our partially
constrained vielbein formulation is such that only the boost part of the
Lorentz transformations propagates so that its perturbation variables
can be written as
\begin{equation}
\delta\omega_{0I} = \partial_I v +v_I\,,
\label{eq:decompv}
\end{equation}
We also need to consider the perturbation of the quasidilaton field $\sigma$ as
\begin{equation}
\sigma = \sigma(t)+\mpl \delta\sigma \,.
\end{equation}
All the perturbed quantities have both time and space dependence. Our perturbations on top of the de Sitter attractor neglecting the standard matter field contain na\"ively counted fourteen degrees of freedom (dof), where two of them are the massive transverse--traceless symmetric spatial tensor fields ($h_{ij}$), six of them are divergence-free spatial vector fields ($B_i$, $E_i$, $v_i$) and the remaining six dof are scalar fields ($\Phi$, $B$, $\psi$, $E$, $v$, $\sigma$). We will see explicitly that the vector modes $B_i$ and $v_i$ are actually non-dynamical and the same is true for the scalar fields $\Phi$, $B$ and $v$.

Throughout this Section, the background is the de Sitter attractor; the quantities $r$, $H$, $X$, $N_\eff$ and $a_\eff/a$ have constant values dictated by the background equations of motion discussed in the previous Section.

Let us first start our analysis of the tensor perturbations. For this we decompose the tensor field $h_{ij}$ in Fourier modes with respect to the spatial coordinates
\begin{equation}\label{tensor_Fourier}
h_{ij}=\int \frac{\d^3k}{(2\pi)^{3/2}}h_{ij}(\vec{k},t) \,{\rm e}^{i\vec{k}\cdot\vec{x}} 
\,,
\end{equation}
then apply the above perturbation decomposition to our Lagrangian and expand it to quadratic order in the tensor perturbations. On the de Sitter attractor solution, the quadratic action becomes
\begin{equation} 
S^{(2)}_{\rm tensor} = \frac{\mpl^2}{8}\int
  d^3k\,dt\,a^3\,\left[\dot{h}_{ij}^\star
    \dot{h}^{ij}-\left(\frac{k^2}{a^2}+m_{T}^2\right)h_{ij}^\star
    h^{ij} \right]\,,
\end{equation}
where it should be understood that $h_{ij}$ is the mode function in momentum space. The mass of the tensor graviton is a constant and is given by
\begin{eqnarray}
m_T^2 &=& m^2\Gamma -\frac{\alpha\,\beta\,\omega\,a_\eff\,X\,H^2}{2\,a\,N_\eff}\,,
\end{eqnarray}
where $\Gamma$ was defined in Eq.~(\ref{eq:Xfunctionsbeta}).
One immediate observation is that the tensor perturbations behave exactly as in the metric formulation, on the late-time de Sitter background. This is due to the fact that the boost parameters contribute only to the vector and scalar perturbations. The tensor perturbations do not yield any ghost nor gradient instabilities as usual for variants of dRGT massive gravity, while the absence of a tachyonic instability can be ensured if $m_T^2>0$.

As we did for the tensor perturbations, we first decompose the vector modes $E_i$, $B_i$ and $v_i$ in Fourier modes. The action for the vector modes around the de Sitter attractor yields
\begin{eqnarray}
S^{(2)}_{\rm vector}&=& \frac{\mpl^2}{16}\int d^3k\, dt\, k^2 a^3 \left[\dot{E}_i^\star \dot{E}^{i}-\frac{2}{a}\,\left(\dot{E}_i^\star B^i + B_i^\star \dot{E}^i\right)-m_T^2\,E_i^\star E^i+\frac{4}{a^2}\,B_i^\star B^i
\right.\nonumber\\
&&\left.\qquad\qquad\qquad-\frac{8\,\alpha\,\omega\,a_\eff^2 H^2\,\left(2+r-\frac{a\,N_\eff}{a_\eff}\right)}{k^2(r-1)a^2 N_\eff}v_i^\star v^i+\frac{8\,\alpha\,\omega\,a_\eff^2H^2}{k^2(r-1)a^2 N_\eff}\left(B_i^\star v^i+v_i^\star B^i\right)
\right]\,.
\label{eq:cvf-vector-initial}
\end{eqnarray}
Solving for the non-dynamical degrees results in
\begin{equation}
v_i = \left(2+r-\frac{a\,N_\eff}{a_\eff}\right)^{-1}\,B_i\,,\qquad
B_i = \frac{a}{2}\,\left(1+\frac{2\,\alpha\,\omega\,a_\eff^2H^2}{k^2(r-1)\,N_\eff\left[2+r-\frac{a\,N_\eff}{a_\eff}\right]}\right)^{-1}\dot{E}_i\,.
\end{equation}
Putting these expressions back into the action, we obtain 
\begin{equation}
S^{(2)}_{\rm vector}= \frac{\mpl^2}{16}\int d^3k\, dt\, k^2 a^3 \left[K_V^2 \dot{E}_i^\star \dot{E}^{i}-m_T^2\,E_i^\star E^i
\right]\,,
\end{equation}
with the prefactor of the kinetic term given by
\begin{equation}
K_V^2 \equiv \left[1+\frac{k^2 N_\eff(r-1)\left(2+r-\frac{a\,N_\eff}{a_\eff}\right)}{2\,H^2\alpha\,\omega\,a_\eff^2}\right]^{-1}\,.
\end{equation}
For the absence of the ghost instability one has to simply impose $K_V>0$. The absence of the tachyonic instability in the tensor sector equally means the absence of gradient instabilities in the vector modes.

Finally, we move on to the scalar perturbations. As mentioned above not all of the scalar dof propagate. In fact, we can use the
equations of motion for the non-dynamical degrees $B$, $\Phi$ and $v$ to integrate them out
\begin{equation}
\left(\dot{\psi}+\frac{k^2}{6}\,\dot{E}\right)-H \,\Phi +\frac{H^2\alpha\,\omega\,a_\eff^2}{2\,a^2 N_\eff}\,\frac{\delta\sigma}{H} - \frac{H^2\alpha\,\omega\,a_\eff^2}{2(r-1)a\,N_\eff}\,v=0\,,
\label{eq:cvf-scaeq1}
\end{equation}
\begin{eqnarray}
&&H\left(\frac{k^2 B}{a}-3\,H \Phi+3\,\dot{\psi}\right)+\frac{3\,H^2\alpha\,\omega\,a_\eff^2}{2(r-1)a^2 N_\eff}(\psi-\delta\sigma)+\frac{H^2\alpha\,\omega\,a_\eff^3}{2\,a^3N_\eff^3}\left(\alpha\,\Phi+\beta\,r X\,\delta\sigma\right)
\nonumber\\
&&
\qquad\qquad\qquad\qquad\qquad\qquad\qquad\qquad\qquad+\frac{k^2}{a^2}\left(\psi+\frac{k^2}{6}E\right)
-\frac{H\alpha\,\omega\,a_\eff^3}{2\,a^3N_\eff^2}\,\delta\dot{\sigma}=0\,,
\label{eq:cvf-scaeq2}
\end{eqnarray}
\begin{equation}
\frac{H^2k^2\alpha\,\omega\,a_\eff^2}{2(r-1)a^2 N_\eff^2}\,\left[B-\frac{(r-1)^2X\,\beta}{a_\eff H}\,\delta\sigma-\left(2+r-\frac{a\,N_\eff}{a_\eff}\right)v\right]=0\,.
\label{eq:cvf-scaeq3}
\end{equation}

After solving these equations and using them back in the action
quadratic in scalar perturbations, we still have three degrees of
freedom: $\psi$, $\delta\sigma$ and $E$. On performing the following field redefinitions
\begin{equation}
\delta\sigma=Y_1 +\frac{\alpha}{\alpha+\beta r X}\,\left(\psi + \frac{k^2}{3}\,Y_2\right)\,,\qquad
E=2Y_2\,,
\label{eq:fieldredef}
\end{equation}
the mode $\psi$ becomes a Lagrange multiplier (removing the would-be
Boulware-Deser ghost). After integrating it out, we have schematically
the following action:
\begin{equation}
S^{(2)}_{\rm scalar}=\frac{\mpl^2}{2}\int dt\,a^3 d^3k\left[ \dot{Y}^\dagger\cdot K\cdot \dot{Y}+\dot{Y}^\dagger\cdot \mathcal{M}\cdot Y - Y^\dagger\cdot \mathcal{M}\cdot \dot{Y}-Y^\dagger \cdot \Omega^2\cdot Y\right]\,,
\end{equation}
where $K$ and $\Omega^2$ are symmetric $2\times2$ matrices, while $\cal M$ is an anti-symmetric $2\times2$ matrix. The components of these matrices are not suitable for presentation. On the other hand, all we are interested in are the positivity of the kinetic terms and the positivity of the coefficients of the $k^2$ terms in the dispersion relations of the eigenfrequencies. Both of these tasks become practical 
in the superhorizon limit, i.e.\ $k \gg a\,H$. 

From the positivity of the kinetic matrix $K$ we can read
out the no-ghost conditions as $\text{NG}_1 > 0$ and $\text{NG}_2>0$,
where
\begin{eqnarray}
\text{NG}_1 &\equiv& K_{11} =
 \frac{\omega\,a_\eff^3}{a^3 N_\eff}\,.
\label{NG1vielbein}
\\
\text{NG}_2 &\equiv& \frac{\det K}{K_{11}}=
\frac{6 H^2 m_\text{T}^2 \beta^2 X^2 a^6}{a_\eff^2}
+
\frac{3H^4 \omega\, \alpha\, a^5 
}{2 a_\eff N_\eff^2 (r-1)^2}
\left\{
\alpha\left[
\omega \Bigl(\frac{a_\eff}{a}\Bigr)^5
- 6 N_\eff^3
\right]
-6 \beta (r-1) X\,
N_\eff \frac{a_\eff^2}{a^2}
\right\}.
\label{NG2vielbein}
\end{eqnarray}
The counterparts of the no-ghost conditions $\text{NG}_1>0$ and $\text{NG}_2>0$ in the metric formalism are given in 
Eqs.~(\ref{eq:metric-kappas})--(\ref{eq:metric-noghost}).  Indeed the condition $\text{NG}_1 >0$
is equivalent to the first condition in (\ref{eq:metric-noghost}).
The other condition $\text{NG}_2>0$, however, does not reduce to the second of (\ref{eq:metric-noghost}). This difference is one
of the new features introduced by switching from the metric formalism
to the partially-constrained vielbein formalism. 

Finally, the squared sound speeds $c_s^2$ can be obtained by solving the quadratic equation
\begin{equation}
\frac{(\text{NG}_1) \, (\text{NG}_2)}{a^4} \, (c_s^2)^2 - A \, c_s^2 + B = 0,
\end{equation}
whose coefficients can be expressed as
\begin{align}
A&\equiv
\frac{4   \beta ^2 m_T^4 X\,^2 (r\,-1) \bigl[ (r\,+2)\ta -N_\eff  \bigr] }{\alpha \ta^2 }
\notag \\
&\quad
+ \frac{2 H\,^2 m_T^2 \omega }{\ta N_\eff^2 (r\,-1)^2}
\left\{
\omega \alpha  \ta^6 (r\,-1) 
\right.
\notag \\
&\qquad\qquad\qquad\quad
\left.
+N_\eff  \left[
 \left(8 r\,^2+r\,-7\right)\ta^4 
+ (-8r\,^2-9r\,+13)\ta^3 N_\eff 
- (4 r\,+3)\ta^2 N_\eff^2
+6    (r\,+1)\ta N_\eff^3
-3 N_\eff^4
\right]
\right\}
\notag \\
&\quad
+\frac{ \ta^2 \alpha H\,^4 \omega^2 }{2 N_\eff^3 (r\,-1)^3}
\left\{
-    \omega \alpha  (r\,-1)\ta^3  \left(\ta^2-3 N_\eff^2\right)
\right.
\notag\\
&\qquad\qquad
\left.
+2 N_\eff  \left[
(3 + r\, - 20 r\,^2 + 16 r\,^3)\ta^3
- (r\,-1) (16 r\,-7)\ta^2 N_\eff  
-3  (3 - 9 r\, + 8 r\,^2)\ta N_\eff^2
+3  (5 r\,-3)N_\eff^3
\right]
\right\}\,,\nonumber
\\
B&\equiv
-\frac{4   m_T^4 N_\eff^2 \left[ (3 r\,-2)\ta +(1-2r\,) N_\eff  \right] }{\alpha \ta^2 (r\,-1)}
+ \frac{2  H\,^2 m_T^2 \omega }{ (r\,-1)^2}
\left[
 \omega  
\alpha (r\,-1) 
\ta^3
+ (-7 + 13 r\, - 8 r\,^2)N_\eff  \ta
+(1+r\,)N_\eff^2 
\right]
\notag \\
& \quad
-\frac{ \ta^2 \alpha H\,^4 \omega ^2 }{2 N_\eff  (r\,-1)^3}
\left[
\omega \alpha  (r\,-1)  \ta^3
-2 (3 - 8 r\, + 4 r\,^2) N_\eff  \ta
+2 (r\,-2)N_\eff^2 
\right]~,
\end{align}
where $\ta \equiv a_\eff/a$. In order to have real $c_s^2$, one has to satisfy
\begin{equation}
A^2 \ge \frac{4\,B\,(\text{NG}_1)(\text{NG}_2)}{a^4}\,,
\end{equation}
while the positivity of the squared sound speeds, necessary to avoid gradient instability, requires $A>0$ and $B>0$. Again there is a qualitative difference compared to the metric formalism. In the metric formalism, where the propagation speeds are given by Eq.~(\ref{eq:metric-cs2}), the gradient term in the dispersion relation of one of the modes is zero, i.e.\ there is a non-propagating mode. However, we see that the situation has now changed in the constrained vielbein formulation, as the said mode acquires a non-zero sound speed.

\section{Summary and discussion}
\label{Sec:summary}

In this work we studied the new quasidilaton theory in the partially-constrained vielbein formulation. This formalism mimics the ideas proposed and studied recently in \cite{DeFelice:2015hla,DeFelice:2015moy}, where it was aimed to remove the unwanted and unstable degrees of freedom of the dRGT theory and keep only the tensor modes. Even if in our formalism the vector and scalar modes are kept as well, we get rid of the BD ghost at the fully non-linear level. It is well known that one can decompose any general vielbein into a Lorentz boost and rotation of a triangular vielbein. Using advantage of this fact, in \cite{Hinterbichler:2015yaa} it was shown that the integration of the boost parameters results in a linear Hamiltonian. Notwithstanding the integration of the remaining rotation parameters gives a Hamiltonian highly non-linear in the lapses \cite{deRham:2015cha}. Hence, we constructed our partially constrained vielbeins exactly with the purpose that the rotation parameters can no longer reintroduce the non-linearities in the lapses. Our aim in the present work was to examine the stability
of cosmological solutions in the new quasidilaton theory in this new formulation.  In the metric formalism,
the BD ghost may appear above an energy scale that depends on the
parameter $\beta$ in the quasidilaton coupling, and more specifically
$\beta$ must be tuned sufficiently small to make the mass of the BD
ghost larger than the cutoff scale of the theory as argued in the
summary of Ref.~\cite{Mukohyama:2014rca}.  In the
partially-constrained vielbein formalism, the BD ghost is absent
nonlinearly and hence the fine-tuning mentioned above becomes
unnecessary.  Thanks to this property the allowed parameter region is
greatly enlarged in the latter formulation.

The change in the formalism does not affect the background solutions and also perturbations in the tensor and vector sectors.  Since we gave the detailed calculation of the perturbations in the metric formalism in the appendix, they can be directly compared with those in the partially-constrained vielbein formalism. As usual the ghost and gradient stability of the tensor perturbations is guaranteed. Additionally, by imposing the mass of the tensor modes to be positive we avoid tachyonic instability. On the other hand the stability of the vector perturbations is granted only by further demanding $K_V>0$. The scalar perturbations yield the new difference between the different formulations. The constraint for the absence of ghost and gradient instability is crucially changed.

The partially-constrained vielbein formalism can be applied also to
the bimetric theories to modify stability properties of cosmological
solutions.  It would be interesting to examine implications of such
modifications to the cosmology in various extensions of massive
gravity theories.

\begin{acknowledgments}
The work of AEG is supported by STFC Consolidated Grant ST/L00044X/1. L.H. acknowledges financial support from Dr. Max R\"ossler, the Walter Haefner Foundation and the ETH Zurich Foundation. The work of S.M. was supported in part by Grant-in-Aid for Scientific Research 24540256 and World Premier International Research Center Initiative (WPI), Ministry of Education, Culture, Sports, Science and Technology (MEXT), Japan. Part of his work has been done within the Labex ILP (reference ANR-10-LABX-63) part of the Idex SUPER, and received financial state aid managed by the Agence Nationale de la Recherche, as part of the programme Investissements d'avenir under the reference ANR-11-IDEX-0004-02. He is thankful to colleagues at Institut Astrophysique de Paris, especially Jean-Philippe Uzan, for warm hospitality. N.T.\ was supported by the European Research Council grant no.\ ERC-2011-StG 279363-HiDGR.
\end{acknowledgments}

%%%%%%%%%%%%%%%%%%%%%%%%%%%%%%%%%%%%%%%%%%%%%%%%%%%%%%%%%%%%

%%%%%%%%%%%%%%%%%%%%%%%%%%%%%%%%%%%%%%%%%%%%%%%%%%%%%%%%%%%%

\appendix

\section{Metric formulation of the new quasidilaton theory}
\label{Sec:metric}

\label{sec:quasidilatondRGT}

In this appendix we summarize the computation of cosmological perturbations around de Sitter background for new quasidilaton theory in the metric formulation. Although this analysis has been already worked out in Ref.~\cite{Mukohyama:2014rca}, to facilitate direct comparison with the partially-constrained vielbein formalism, we include some intermediate steps to trace the difference between the two approaches. The action we consider is
\begin{equation}\label{action_NQD_effcoupl}
\mathcal{S} = \int \mathrm{d}^4x \left\{ \frac{\mpl^2}{2} \sqrt{-g}\left[R[g]-2\Lambda+2m^2(\alpha_1\mathcal{U}_1+\alpha_2\mathcal{U}_2+\alpha_3 \mathcal{U}_3+\alpha_4 \mathcal{U}_4) \right]-\frac{\omega}{2}\,\sqrt{-g_{\rm eff}}\,g_\eff^{\mu\nu}\partial_\mu\sigma\,\partial_\nu\sigma \right\}\,,
\end{equation}
where 
\begin{eqnarray}
\mathcal{U}_1[\mathcal{K}] &=& [\K]\,, \nonumber\\
\mathcal{U}_2[\mathcal{K}] &=& \frac{1}{2}\left( [\K]^2-[\K^2]\right)\,, \nonumber\\
\mathcal{U}_3[\mathcal{K}] &=& \frac{1}{6}\left( [\K]^3-3[\K][\K^2]+2[\K^3]\right)\,,  \nonumber\\
\mathcal{U}_3[\mathcal{K}] &=& \frac{1}{24}\left([\K]^4-6[\K]^2[\K^2]+3[\K^2]^2+8[\K][\K^3]-6[\K^4]\right)\,,
\end{eqnarray}
and the effective metric is given by~\cite{Mukohyama:2014rca}
\begin{equation}
g_{\mu\nu}^{\rm eff}=\alpha^2 g_{\mu\nu}+\beta^2 \,{\rm e}^{2\sigma/\mpl}f_{\mu\nu}+2\,\alpha\,\beta \,{\rm e}^{\sigma/\mpl} g_{\mu\rho}\left( \sqrt{g^{-1}f}\right)^\rho_\nu\, .
\end{equation}
Similarly, the building block tensor $\K$ of the original dRGT theory of massive gravity is modified into
\begin{equation}
\K^\mu _\nu[g,f] =\delta^\mu_\nu -\,{\rm e}^{\sigma/M_{\rm Pl}} \left(\sqrt{g^{-1}f}\right)^\mu_\nu \,,
\end{equation}
with the presence of the $\sigma$ field, while the reference metric $f$ is kept the same
\begin{equation}\label{Stueckelbergfields}
f_{\mu\nu} = \eta_{ab}\partial_\mu \phi^a \partial_\nu\phi^b\,,
\end{equation}
with the St\"uckelberg fields $\phi^a$. Note that there is no disformal transformation to the fiducial metric anymore as in the extended quasidilaton scenario \cite{DeFelice:2013tsa}, i.e.\ here the field space is 4-dimensional. The purpose of introducing this disformal factor was actually to render the self-accelerating late-time asymptotic solutions stable. In \cite{Mukohyama:2014rca} it was shown that this purpose is achieved also with the coupling to the effective metric (although it is still compatible with the global quasi-dilaton symmetry).

We concretize our dynamical background metric to be of the homogeneous and isotropic flat FLRW form
\begin{equation}
ds_g^2=-N^2 dt^2 +a^2 \delta_{ij} dx^idx^j\,.
\end{equation}
We then choose the unitary gauge, i.e.\ $\phi^0=\varphi(t)$, $\phi^a=a_0\,x^a$ giving the fiducial metric
\begin{equation}
ds_f^2= f_{\mu\nu}dx^\mu dx^\nu= -M^2 dt^2 +  a_0^2\delta_{ij} dx^idx^j\,,
\end{equation}
where $M=\dot{\varphi}$. Finally, for a homogeneous background of quasi-dilaton $\sigma(t)$, the action (\ref{action_NQD_effcoupl}) takes the following form:
\begin{equation}
\frac{S}{V} = \mpl^2\int dt \,a^3 N\,\left[-\Lambda-3H^2-m^2\left(\rho_{m,g}+r\,X^4\rho_{m,f}\right)+\frac{\omega\,a_\eff^3\,\dot{\sigma}^2}{2\,\mpl^2a^3N_\eff N}\right]\,,
\label{eq:minisuperspace}
\end{equation}
where now we have $r\equiv \frac{M\,a}{a_0\,N}=\frac{\dot{\varphi}\,a}{a_0\,N}$. In the above, for convenience, we used the function
\begin{equation}
U(X)=4\,(X-1)\alpha_1-6\,(X-1)^2\alpha_2+4\,(X-1)^3\alpha_3-(X-1)^4\alpha_4\,,
\end{equation}
which we plugged in Eq.~(\ref{eq:Xfunctionsbeta}) to define the quantities $\rho_{m,g}$, $\rho_{m,f}$, $J$ and $\Gamma$, replacing the $\alpha_n$ coefficients.

At this point, we stress that the action (\ref{action_NQD_effcoupl}) is almost the same as the partially-constrained vielbein action (\ref{eq:action}). The difference between the two formalisms arises from the different choice of quantities that are used in the variation. To be specific, the partially-constrained vielbein formalism can be seen to contain four new auxiliary fields, i.e.\ the boosts. As the choice of cosmological background does not excite these degrees of freedom (due to isotropy), the background of the two formalisms are identical.

The background equations of motion can be calculated simply by varying the action (\ref{eq:minisuperspace}) with respect to the lapse $N$, scale factor $a$, quasi-dilaton field $\sigma$ and the temporal St\"uckelberg field $\varphi$, giving the set of equations listed in (\ref{eq:eomsBG}) for $N=1$.
One of these is a redundant equation, due to the contracted Bianchi identity
\begin{equation}
\frac{\partial}{\partial t} \frac{\delta S}{\delta N} - \frac{\dot{a}}{N}\frac{\delta S}{\delta a}- \frac{\dot{\varphi}}{N}\frac{\delta S}{\delta \varphi}- \frac{\dot{\sigma}}{N}\frac{\delta S}{\delta \sigma} - \frac{\dot{\chi}}{N}\frac{\delta S}{\delta \chi}=0\,.
\label{eq:bianchi}
\end{equation}

In order to compare the stability of the perturbations in the metric formulation with those in the partially constrained vielbein formulation, from here on we will specify to the late time dS attractor detailed in Sec.~\ref{sec:bgsol} and fix the residual gauge freedom in time coordinate by setting $N=1$.
We choose a decomposition that is compatible with Eqs.~(\ref{eq:ADMdecomp-pert})--(\ref{eq:decompe}) at linear order:
\begin{eqnarray}\label{perturbed_metric}
\delta g_{00} &=& -2\,N^2\,\Phi\,,\nonumber\\
\delta g_{0i} &=& N\,a\,\left(\partial_i B+B_i\right)\,,\nonumber\\
\delta g_{ij} &=& a^2 \left[2\,\delta_{ij}\psi +\left(\partial_i\partial_j-\frac{\delta_{ij}}{3}\partial^k\partial_k\right)E+\partial_{(i}E_{j)}\right]\,,
\end{eqnarray}
where the vector perturbations are transverse $\partial^i E_i = \partial^i B_i=0$, and we disregarded the tensor perturbations as they are exactly the same in both formalisms at linear order. We also fix all of the gauge freedom by setting the perturbations for the four St\"uckelberg fields to zero.
Similarly, the quasidilaton $\sigma$ is perturbed as
\begin{equation}
\sigma = \sigma(t)+\mpl \delta\sigma \,.
\end{equation}
Excluding the tensor modes, the action (\ref{action_NQD_effcoupl}) contains na\"ively counted nine degrees of freedom (dof), four of them being divergence-free spatial vector fields ($B_i$, $E_i$). The other five dof are scalar fields ($\Phi$, $B$, $\psi$, $E$, $\delta\sigma$). Of course not all of them are dynamical. In what follows we will investigate the stability conditions of the vector and scalar perturbations above the dynamical background equations after integrating out the non-dynamical degrees of freedom.

We start with the stability conditions of the vector modes and expand the Lagrangian (\ref{action_NQD_effcoupl}) to second order in the vector perturbations:
\begin{equation}
S^{(2)}_{\rm vector} = \frac{\mpl^2}{16}\int d^3k\,dt\,k^2a^3\,
\left[
\dot{E}_{i}^\star\dot{E}^i-\frac{2}{a}\,\left(\dot{E}_i^\star B^i+B_i^\star \dot{E}^i\right) -m_T^2 E_{i}^\star E^i +\frac{4}{a^2}\left(1+\frac{a^2m_T^2}{k^2\,c_V^2}\right)B_i^\star B^i
\right]\,,
\label{metric:action_vectormodes-initial}
\end{equation}
where we defined the following constant:
\begin{equation}
\frac{m_T^2}{c_V^2} \equiv \frac{2\,\alpha\,\omega\,H^2\,a_\eff}{(r+1)^2(r-1)a}\,\left(1+\frac{r\,a_\eff}{a\,N_\eff}\right)\,.
\end{equation}
At this level, there is an immediate departure from the corresponding action in the constrained vielbein formalism~(\ref{eq:cvf-vector-initial}). The latter expression reduces to the metric formulation one if the boost parameter is forced to be $v_i=B_i/(1+r)$~\cite{DeFelice:2015yha}.

We notice that not all of the vector fields are dynamical, indeed, the vector fields $B_i$ do not have any time--kinetic terms. We can therefore compute the equations of motion with respect to $B_i^\star$ and $B_i$ and integrate them out by using the solution
\begin{equation}
B_{i}=\frac{a}{2}\left(1+\frac{a^2m_T^2}{k^2c_V^2}\right)^{-1}\,\dot{E}_i\,,
\end{equation}
after which, the quadratic action in the vector perturbations becomes
\begin{equation}\label{action_vectormodes}
S^{(2)}_{\rm vector} = \frac{\mpl^2}{16}\int d^3k\,dt\,k^2a^3 \,\left(1+\frac{k^2c_V^2}{a^2m_T^2}\right)^{-1}\left[
\dot{E}_{i}^\star\dot{E}^i-\left(m_T^2+c_V^2 \frac{k^2}{a^2}\right) E_{i}^\star E^i 
\right]\,,
\end{equation}
where $c_V^2$ now corresponds to the propagation speed of subhorizon modes.
For the stability of vector perturbations on top of the de Sitter background, we have to impose the right sign for the kinetic and gradient terms. For the absence of ghost instability, we require that the kinetic term is positive. This can be achieved for any $k$ if we impose $m_T^2/c_V^2>0$. 

Last but not least let us concentrate on the stability of the scalar perturbations in the new extended quasi-dilaton massive gravity model with matter field. As we mentioned above, five degrees of freedom appear in form of scalar fields ($\psi$, $\delta\sigma$, $E$, $B$, $\Phi$). We first expand the action (\ref{action_NQD_effcoupl}) to quadratic order in the scalar perturbations in their Fourier modes. We first note that the corresponding kinetic matrix has two vanishing eigenvalues, which signals the existence of two constraint equations which make two out of the five scalar fields non-propagating
\begin{eqnarray}
\mathcal{K}_{\psi, \delta\sigma, E, B, \Phi}=
\begin{pmatrix}
-6&0&0&0&0\\
&
\frac{\omega\,a_\eff^3}{a^3N_\eff}
&0&0 &0\\
&&k^4/6&0&0 \\
&&&0&0 \\
&&&&0
 \end{pmatrix}\,.
\end{eqnarray}
Since the quadratic action does not have any kinetic term for the scalar fields $B$ and $\Phi$, we can compute their equations of motion in order to obtain the corresponding two constraint equations. The equation of motion for $B$ and $\Phi$ are, respectively,
\begin{eqnarray}
&&
\dot{\psi}+\frac{k^2}{6}\,\dot{E}-H\,\Phi+\frac{\alpha\,\omega\,H\,a_\eff}{2\,(r+1)a}\left(1+\frac{a_\eff \,r}{a\,N_\eff}\right)\,\left(\delta\sigma-\frac{a\,H}{r^2-1}\,B\right)=0\,,
\nonumber\\
&&
H\,\left(\frac{k^2\,B}{a}-3\,H\,\Phi+3\,\dot{\psi}\right)+
\frac{H^2\alpha^2\omega\,a_\eff^3}{2\,a^3N_\eff^3}\left[\Phi+\frac{3\,a\,N_\eff^2}{\alpha\,(r-1)a_\eff}\,(\psi-\delta\sigma)\right]
+\frac{k^2}{a^2}\left(\psi+\frac{k^2}{6}\,E\right)
\nonumber\\
&&\qquad\qquad\qquad\qquad\qquad\qquad\qquad\qquad\qquad\qquad\qquad
+\frac{H\,\alpha\,\omega\,a_\eff^3}{2\,a^3N_\eff^2}\,\left[\frac{H\,\beta\,r\,X}{N_\eff}\,\delta\sigma-\delta\dot\sigma\right]=0\,.
\end{eqnarray} 
Comparing these equations with the constrained vielbein formalism counterparts (\ref{eq:cvf-scaeq1})--(\ref{eq:cvf-scaeq3}), we notice that the $\delta\sigma$ equation (\ref{eq:cvf-scaeq2}) remains the same in both formalisms, while the $B$ equation is different. Like the vector modes, the metric formulation can be obtained if the scalar boost parameter $v$ is forced by hand to be $B/(1+r)$. However, in the constrained vielbein formulation, as $v$ is treated as an independent variable, we obtain a completely different perturbation spectrum compared to the metric formulation.

After using the constraint equations, the resulting action depends on the remaining three scalar fields ($\psi$, $\delta\sigma$, $E$). However, looking at the kinetic matrix of these three remaining scalar fields, one immediately observes that it still has a vanishing determinant, meaning that there is a constraint that can be used to integrate out one more scalar degree of freedom. This degree, i.e.\ the would-be BD ghost, becomes manifestly non-dynamical in the field basis (\ref{eq:fieldredef}). Integrating out the now non-dynamical $\psi$, the reduced action takes the following form:
\begin{equation}
S^{(2)}_{\rm scalar}=\frac{\mpl}{2}\int N\,dt\,a^3 d^3k\left[\frac{\dot{Y}^\dagger}{N}\cdot K\cdot \frac{\dot{Y}}{N}+\frac{\dot{Y}^\dagger}{N}\cdot M\cdot Y - Y^\dagger\cdot M\cdot \frac{\dot{Y}}{N}-Y^\dagger \cdot \Omega^2\cdot Y\right]\,,
\end{equation}
where $K$ and $\Omega^2$ are symmetric $2\times2$ matrices, while $M$ is anti-symmetric $2\times2$ matrix. As in the constrained vielbein formulation, these matrices are too bulky for presentation, although we now show their subhorizon limit. 
The kinetic matrix in this limit becomes diagonal at leading order with:
\begin{equation}
K_{11} = \kappa_1 + {\cal O}(k^{-2})\,,\qquad
K_{12} ={\cal O}(k^0)\,,\qquad
K_{22} = \kappa_2 \,k^2+{\cal O}(k^0)\,.
\end{equation}
where
\begin{equation}
\kappa_1 =\frac{\omega\,a_\eff^3}{a^3\,N_\eff}\,,\qquad
\kappa_2 = 
\frac{1}{a N_\eff^2}
\frac{H^2(1-r)X^2 \alpha\,\beta^2\omega a_\eff^3}{(r+1)\alpha^2+(2+3\,r-r^2)X\,\alpha\,\beta+2\,X^2\beta^2}\,.
\label{eq:metric-kappas}
\end{equation}
At leading order in large $k$ expansion, $\kappa_1$ and $\kappa_2 k^2$ also correspond to the eigenvalues of the kinetic matrix $K$. Therefore, the no-ghost conditions for this system in subhorizon scales are simply
\begin{equation}
\kappa_1 > 0\,,\qquad\kappa_2 >0\,.
\label{eq:metric-noghost}
\end{equation}

In order to determine the propagation speeds, we use the fact that the frequency in the UV is dominated by $\omega = c_s\,k + {\cal O}(k^0)$ term. We then solve for the following determinant equation, obtained by considering monochromatic waves in the equation of motion for perturbations:
\begin{equation}
\det\left[-c_s^2 \frac{k^2}{a^2}\,K + (\dot{K}+2\,M+3\,H\,K)\left(-i\,c_s\,\frac{k}{a}\right)+\left(\Omega^2+\dot{M}+3\,H\,M\right)\right]=0\,.
\label{eq:deteq}
\end{equation}
At leading order, the $11$ component of the matrix inside the square brackets goes as $k^2$, the $12$ component goes as $k^2$ and the $22$ component goes as $k^4$. Thus, only the following components actually contribute to the above determinant at leading order:
\begin{align}
&M_{12} = \frac{(r-1)\,\kappa_2\,N_\eff}{2\,H\,a\,a_\eff}\,k^2+{\cal O}(k^0)\,,\nonumber\\
&\Omega^2_{11} = \frac{\kappa_2\,N_\eff^3}{H^2(1-r)\,X^2\,a^4 \alpha\,\beta^2}\left(1+\frac{r\,a_\eff}{a\,N_\eff}\right)\,k^2+{\cal O}(k^0)\,,\qquad
\Omega^2_{12} ={\cal O}(k^{2})\,,\qquad\Omega^2_{22} ={\cal O}(k^{2})\,.
\end{align}
Effectively, at leading order, Eq.~(\ref{eq:deteq}) reduces to
\begin{equation}
-\left(-c_s^2\frac{k^2}{a^2}\,K_{11}+\Omega^2_{11}\right)c_s^2\frac{k^2}{a^2}\,K_{22} - 4\,c_s^2\,\frac{k^2}{a^2}\,(M_{12})^2=0\,,
\end{equation}
whose solutions simply give the following propagation speeds
\begin{equation}
c_{s,I}^2 =\frac{a^2N_\eff^2}{a_\eff^2 }\,,\qquad
c_{s,II}^2 =0\,.
\label{eq:metric-cs2}
\end{equation}
\bibliography{cosmology_BG_NewQuasiDilaton}

\end{document}